# Curating China's Cultural Revolution (1966-1976):
## CR/10 as a Warburgian Memory Atlas and Digital Humanities Interface

**Abstract**

CR/10 is a digital oral history platform that aims to collect and preserve the cultural memories of China's Cultural Revolution (1966-1976). This paper discusses how CR/10 functions as a Warburgian memory atlas and shapes multifaceted narratives of the historical incident. Through ethnographic research and semi-structured interviews with users within and outside academia, I examined the usability of CR/10 among various user groups and proposed design opportunities to further empower the interface. This paper offered a strong case on the datafication of cultural memories among cultural heritage institutions and contributed to digital archiving scholarship with an innovative methodological lens.

## Introduction

Lasting from the decade between 1966 and 1976, China's Cultural Revolution, also named the "Great Proletarian Cultural Revolution" by Mao Zedong and the Chinese Communist Party (CCP), was identified as a necessary move to prevent the return of capitalism to China, to protect the purity of the CCP, and to seek a path for the further development of socialism in China. This "Great Revolution" was called a "cultural" revolution because it started from criticism of the cultural arena and aimed at "touching people to their very soul."[1] Following Mao's death and the eradication of the "Gang of Four" in 1976, the Party's Central Committee officially declared the end of the Cultural Revolution at CCP's Eleventh National General Congress. In June 1981, the Sixth Plenum of the Eleventh Central Committee issued the "Resolution on Certain Questions in the History of Our Party Since the Founding of the People's Republic of China," stating that the Cultural Revolution has "caused the Party, the state, and the people to suffer the most serious setbacks and losses since the founding of the nation;" it was "an event initiated by a leader's mistake and exploited by counter-revolutionary groups, bringing disaster and civil unrest to the party, the state, and the people." Following this statement, elite politicians and intellectuals who were suppressed during the Cultural Revolution regained the power of discourse and started reassessment of the historical incident. Despite the limited access to archival materials of the Cultural Revolution, over the last decades, scholars, artists, and information professionals have made tremendous efforts to rethink and reconstruct image(s) of this historical incident.

40 years after the end of the Cultural Revolution, to promote public remembrance and discussion of this significant period of Chinese histories, the East Asian Library at the University of Pittsburgh launched CR/10 in 2016, a digital oral history project that aims to collect and preserve authentic, distinct experiences and memories of China's Cultural Revolution through 10-minute semi-structured interviews.[2] Using the technique of snowball sampling for interviewee selection, the project has thus far collected more than 300 interviews with ordinary people from different generations, geographies, occupations, and social backgrounds, who have experienced the incident themselves or only learned about it from family, school, digital media, or other circumstances. Each interviewee of the project is given about ten minutes to speak freely of their impressions, memories, and thoughts of the Cultural Revolution. All the interviews were video-recorded and made openly accessible with full transcriptions through the interactive CR/10 website.

CR/10 functions as both a digital archival collection and a digital humanities interface that aims to leverage user-centered interactive design to (1) facilitate a flexible, multi-faceted representation of cultural memories in a digital space, and to (2) engage the general public in the remembrance of this historical incident in the contemporary time. As demonstrated by the creator of the CR/10 project, histories of China's Cultural Revolution are deemed "complex and complicated," as "a person's memory varies according to his or her geographic location, age, profession, family background, and many other factors."[3] The CR/10 website, therefore, aimed to illustrate the complexity of cultural memories through a design approach, enabling space for imagination and intervention among the users.

In this article, I analyze how the CR/10 website functions as a *"Warburgian" memory atlas* to facilitate multifaceted accounts of China's Cultural Revolution. This vision of a Warburgian atlas follows the theories and experiments of Aby Warburg's (1866-1929) and his *Mnemosyne*

*Atlas* project, which demonstrated the rhetorical power of images and their spatial arrangements in soliciting viewers' appreciation and re-evaluation of the antiquity[4]. Through a rhetorical analysis of the CR/10 website and semi-structured interviews with its users both within and outside academia, this paper aims to address two central questions:

*(1) How do the current design features of CR/10 function in constructing multifaceted narratives of China's Cultural Revolution?*

*(2) How are CR/10 received among the users?*

Building upon the research, this article proposes recommendations that can be applied to improve the curation of CR/10 as a *participatory digital humanities interface* that sustains memories of China's Cultural Revolution. With the rapid paradigm shifts in the area of digital archives, characterized by trends such as computational archival science and design thinking, this paper contributes to the Special Issue by (1) offering a specific case on datafication of cultural memories among cultural heritage institutions, and (2) demonstrating how Aby Warburg's intellectual experiment can empower this datafication process for cultural information.

## Literature Review

### The Chinese Cultural Revolution(s) Represented

Memories of China's cultural revolution have been represented in various forms of media. These media channels function as what Michael Lynch and Steve Woolgar later identified as representational devices,[5] which reveal the untraceable cultural memories among different social groups. Memoirs is one of the major media channels where memories of China's Cultural Revolution are embodied and expressed. Most of the memoirs available in both Chinese and English provide accounts on traumatic experiences during China's Cultural Revolution from the perspective of intellectuals, Sent-down Youths, and apologetic Red Guards who survived the Cultural Revolution.[6] As Margo Gewurtz stated, these groups eagerly acted as the "carriers of the word" (*daiyan ren*) for those who moved "from the culturally dispossessed below to the wielders of political power above," and fought with their pen for the "recovery of memories."[7] Memoirs, in this case, serve as a major genre for the intellectuals and assumably the "victims" group of the Cultural Revolution to give vent to their traumatic experiences, and therefore, displays a chaotic past of the Cultural Revolution. Oral history testimonies in China, in comparison, go beyond the narratives of intellectual elites currently in power to also include voices of ordinary people on the historical incident.[8] Despite the genuine intention, however, historical accounts of this kind have also been heavily impacted by the time, geography, as well as the cultural and societal situations of the speaker[9].

Compared with the victim accounts and the socialist speeches[10] pervasive in the genre of memoir and oral history testimonies, literary works and cinematic treatments of the topic have demonstrated a more complex, critical, and reflective attitude towards the Cultural Revolution.[11] Some works even expressed nostalgic and passionate emotions towards the Cultural Revolution, by means of featuring happy childhoods and the "love stories and simple dramas of everyday lives" during the Cultural Revolution.[12]

With the rise of social media, cyberspace has also become one of the major venues to publicly memorize China's Cultural Revolution. According to Thomas Heberer and Feng-Mei Heberer's research on discussions of Cultural Revolution in the Internet and social media platforms (e.g., Weibo, the major microblogging site in China), the online engagement creates a "new participatory culture" and "bottom-up discourse" of the Cultural Revolution, which enabled more diversified opinions to be heard.[13] The post-80's generation become one of the leading forces in the cyberspace reconstruct social memories of China's Cultural Revolution.[14]

This non-exhaustive survey of memory works in various media forms have strongly suggested the revival of Cultural Revolution memories as a collection of group narratives. Exposure to one specific group narrative would only unveil a will-saturated snapshot of the past, rather than a complete historical picture.

***Datafication of Cultural Revolution Memories***

As the complex relationships between massive representations and group narratives unfold, it becomes increasingly important to collect and preserve the numerous memory data. With the rise of digitization of cultural records, cultural heritage institutions across the globe have begun to preserve and "open up" the Cultural Revolution memory works in widely accessible digital forms. Their engagement has concentrated on four major categories: databases, digital archive collections, LibGuides, and online exhibitions. To build databases for scholarly research, librarians and archivists have made efforts in both collecting databases on this subject that are already available in China and creating new ones to better serve research demands of their patrons. The first attempt can be exemplified by the new Chinese Cultural Revolution Database at Yale University[15] and the University of California, Santa Barbara Libraries[16]; while the second has been illustrated by the newly developed series of contemporary political databases by East Asian libraries at multiple U.S. universities[17]. Digital archive collections focusing on this subject have also been developed worldwide, including examples such as the *Chinese Cultural Revolution (1966-1976)* at the Wilson Center,[18] *Chinese Digital Archive 1966-1976* at the Asia Portal of the Nordic Institute of Asian Studies,[19] and the *Chinese Digital Archive 1966-1976* at the Australian National University.[20] To better make use of the massive resources and engage wider audiences, librarians and curators developed online exhibitions showcasing new archival materials or unique collections[21], in addition to the more conventional LibGuides approach to collections[22].

Advancement in digital technologies has pushed forward paradigms in datafication of cultural records. The recent rise of computational archival science (CAS) as a trans-disciplinary research and practice field advocates the "application of computational methods and resources to large-scale records/archives processing, analysis, storage, long-term preservation, and access," in order to "improv[e] efficiency, productivity and precision in support of appraisal, arrangement and description, preservation, and access decisions".[23] The CAS paradigm argues for an integration of computational methods with archival conceptualizations, theories, and practices, which essentially has raised a demand for digital archival programs to not just store and preserve, but also to *function* for further computational data analysis or broader engagement with data in other forms. Emerging cases have illustrated that computational tools can be leveraged to increase usability of archival materials and that digital archives designed with computational thinking can facilitate data analyses among users.[24]

This increasing demand for better interactability of digital archives is also echoed in works on interface design from the perspectives of human-computer interaction and digital humanities. An early outlook for interface design focused on the engineering details to achieve optimized effects and efficiency of functional *machines*. Starting around the 1960s, the emphasis on human-computer interaction raised the needs to design interface adjustable for user needs, which aimed at creating an *empty space* between machines and users.[25] Scholars held that an interface should demonstrate the *power of intervention*, able to influence user behaviors during the user-machine interactions, and mediate intellectual and cognitive activities.[26] Johanna Drucker further argued that interface should be a dynamic zone in which reading takes place, and called for a *humanistic approach* to interface design. An interface following this design principle does not only aim to increase usability and efficiency, but also looks at the function of the interface in engaging interpretations and its ability of being critical.[27] Visual data displays incorporated in the interface, in particular, were demonstrated as an important component to achieve the humanistic interface design.[28] Analyzing a number of historical digital humanities interfaces (e.g., the Valley of the Shadow project), Claire Warwick's has also illustrated the rhetorical power of interface design in shaping historical accounts, memories, and scholarship.[29]

As a case where interface design is used as a method to facilitate public engagement with Cultural Revolution memories, CR/10 stands out from the previous archival attempts in various ways. First, CR/10 collected multiple perspectives and narratives of the Chinese Cultural Revolution by recruiting interviewees across various social groups (e.g., gender, occupation, age, economic status). Second, such multi-narratives are visually organized and displayed through the interactive platform of CR/10, particularly the features of timeline and map, which encourages continuous public interventions in the memories shaping process. As will be discussed further in the following sections, the collection of multi-narratives and the digital mapping of them transformed CR/10 into a *memory atlas*, building upon the legacy of Aby Warburg.

### Aby Warburg and the Mnemosyne Atlas

Aby Warburg (1866-1929) was a German art historian and cultural theorist. As an art historian, Warburg's work focused on the legacy of the classical world and the Renaissance period[30]. But more broadly speaking, Warburg's scholarship also attempted to address the relationships between visual objects, techniques and strategies, and their impacts on shaping understandings or delivering knowledge[31]. Warburg proposed the metaphor of *image vehicle* (*Bilderfahrzeuge*), where he demonstrated that images bear the power to transmit and mobilize ideas and cultural memories to travel across time and space.[32]

*Mnemosyne Atlas* was one of his experimental attempts to illustrate how images and the spatial arrangement of them create rhetorical powers in (re)shaping memories and interpretation of the antiquity[33]. The *Mnemosyne Atlas* project started in December 1927 and remained unfinished until Warburg's death in 1929, where Warburg used 40 wooden panels to piece together around 1,000 visual materials of all kinds, including pictures, maps, calendars, and diagrams collected from books, magazines, or newspapers. The purpose of this project was to compose a pictorial atlas that demonstrate how ideas and memories of the past sustained into the present. Themes in the *Mnemosyne Atlas* included "coordinates of memory," "vehicles of tradition," "the classical tradition today," and "archaeological models." No captions or texts are accompanied with the panels to elaborate on their meaning, purposes, or implications. Among the forty panels, panels A-C serve as the preliminary guide to the project, where Warburg showcased a certain number of *pathways* to navigate each panel and laid out essential *grammars* (or *syntax*) for the project[34].

The programmatic design of the *Mnemosyne Atlas* followed a central principle of *distance.* According to Warburg's own introductions to the *Mnemosyne Atlas*,

> "[the] conscious creation of distance between the self and the external world may be called the fundamental act of civilization. Where this in-between space gives rise to artistic creativity, this awareness of distance can achieve a lasting social function – on whose success or failure as an instrument of mental orientation the fate of culture depends."[35]

In Warburg's own accounts, the creation and disruption of distance is the essential mechanism of meaning production for the *Mnemosyne Atlas*, which can be more specifically characterized from three aspects. First is a sense of distance in physical space, which refers to the distance between the photographic images and the narrative gap it produces. An image or pictorial reproduction is placed together with another because they are related in some ways; but the specific relation(s) among them have never been articulated. As David Marshall illustrated, "each image table can be understood as a series of potential exercises in which we move between stances."[36] From a viewer's perspective, the meaning of a panel exists not only in image motifs, but more importantly, in the spatial arrangements of these visual objects. Such a notion of distance was also echoed in later concepts such as the *split-screen visualization display* by Michael Lynch[37] and the *combination diagram* by Scott Montgomery[38], influencing a broader field of visualization research.

A second distinct feature of the *Mnemosyne Atlas* is its *duality* between the past and contemporary time – a critical, constant reflection on the current situation of both the creator and the viewer.[39] *Duality* acknowledged the gap between the visual objects and the viewer, through which the act of interpretation takes place. All the images and motifs used to create the *Mnemosyne Atlas* were taken out of their original contexts of production and appreciated in the viewers' current situation. This situation includes both the larger temporal context where the viewer positions themselves and the individual state – be it mental, social, or cultural.

Last but not least, a Warburgian sense of a distance was also characterized with the *non-linear discourses* the *Mnemosyne Atlas* created, particularly with instrumental assistance from pathways.[40] Compared with the book tradition that emphasizes the creation of fixed, logical narrative flows, the *Mnemosyne Atlas* strived to make arguments that were neither finalized nor single. Such an attempt in physical space, however, proved to be challenging. As Warburg recalled the difficulties in his diary,

> "The regrouping of the photo-plates is tedious: How to show the struggle for an antiquizing ideal style as 1) an argument between no[rth] and so[uth] and 2) as arguing gesture? […] mass displacement within the photo plates. […] Difficulty: the placement of Duccio. […] Pushing around frames with Freund. […] The arrangement of plates in the hall causes unforeseen inner difficulties. […] Begun to cut out all the gods."[41]

The rise of information technologies has transformed Warburg's experiment into digital space and made it easier to deliver research outcomes in the way that Warburg had imagined in the 1920s. Martin Warnke and Lisa Dieckmann presented the project of *Meta-Image*, where they implemented Warburg's methodological approach in *Mnemosyne Atlas* in the design of the interactive platform.[42] With this research platform, scholars working with visual objects may replicate the same procedures and steps that Warburg had advocated for. It is a strong case demonstrating the possibilities of remodeling the *Mnemosyne Atlas* in the digital space. For this study, Aby Warburg's intellectual experiment offered a conceptual model and specific approaches to build upon what Drucker conceptualized as s humanistic interface, and transform a digital archival collection into an interactive, narrative-shaping memory atlas of China's Cultural Revolution.

## Curation of CR/10 as a Warburgian Memory Atlas

CR/10 website functions as a digital Warburgian memory atlas from four dimensions. Primarily, CR/10 is organized based upon fragmented, independent oral history video units that create the Warburgian mechanism of distance. Second, CR/10 aims to be a moving gallery in the digital space that invites audiences to interact with the videos, rethink the historical incident, and create interpretations from their own perspectives. Third, this project leverages specific design features as grammars and syntaxes to facilitate the construction of memory narratives among the users. And finally, CR/10 generates non-linear and multifaceted narratives of the historical incident enabled by navigational pathways on the website.

### *Datafication of Fragmented Memories*

CR/10 started with the conversion of historical human experiences into digital data. As illustrated in fig.1, the memories collected through oral history interviews of CR/10 project were transformed into open access data through four major steps. The first step was video recording of all interviews. The recording process aimed to preserve as much information as possible, including not only the interviewees' verbal accounts, but also their body languages, face expressions, interactions with the interviewer, and the outside environment. Following the video collection, the second stage of datafication focused on processing of all videos. Due to the political sensitivity of China's Cultural Revolution as a historical subject, curators of the project went through a de-identification process, covering some of the interviewees' images or distorted their voices to protect privacy and identities of individuals. The third step for datafication was the transcription of interview content, creating textual data that can be used later to highlight essential messages from the original videos. The transcription data set up a basis for the metadata creation, which was the final stage. Metadata for this collection followed the rules of Dublin Core and contained essential information of the materials, such as the title, creator, identifier, date of creation, material type, specific descriptions, and rights information. The "description" field reveals detailed information about the interviewer that cannot be classified into other categories, e.g., the period when the interviewee was born, geographical areas where they lived during the Cultural Revolution, their family background, occupation, and the highest level of education. Information of this kind, as I will discuss further in the following sections, can be useful to empower the CR/10 website design, especially to increase its searchability as a digital humanities interface.

With the memory data at hand, curation of the CR/10 website to display them went through three major phases. In the first phase, 32 processed interviews were demoed for open access, which created a prototype and initial blueprint for the website. Proceeding from this stage in 2017, a team of China Studies experts, librarians, archivists, and graduate students outlined the basic framework and crucial features of the website, while populating its content with materials such as an introductory video to the project and a trailer of the interviews. The final stage was devoted to expanding on the interviews, advocating for the project, and refining the website. The first version of the website online contained the initial batch of 67 interviews; and currently, the number of included videos on the website has exceeded 300. Key design features of CR/10 website that shaped it as a Warburgian memory atlas, including the timeline, map, and what I would like to call the "ten-minute grammar," are discussed below.

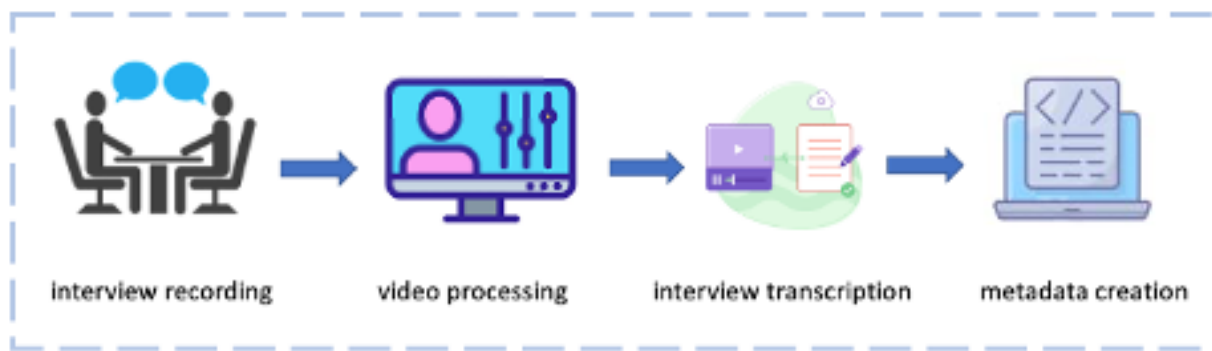

**Fig. 1.** Oral history data processing stages in CR/10.

*The Ten-Minute Grammar*

The ten-minute grammar refers to the principle that each interviewee has only ten minutes to speak of their experiences or memories of the Cultural Revolution. With this setting, each interviewee is expected to focus on the experiences, events, or ramifications that they would most want to reveal. This ten-minute grammar was created based on considerations from three aspects. The primary purpose of the ten-minute grammar was to record the most immediate responses from the interviewees. Immediate responses usually reveal an individual's most unforgettable memories of the historical period. The ten-minute grammar, then, filters an interviewee' memory bank and selects the most unique snippets from them. Second, the ten-minute grammar was inspired by the user-centered design principle. Ten minutes may be as long as we can pay attention to a single video, given the decreasing attention span of people among information distractions.[43] Focusing on the ten-minute grammar assists audiences to grasp central messages of a video and maximizes the potential of CR/10 as an effective teaching and learning platform. The final consideration for the ten-minute grammar was mostly practical. For individuals who experienced the Cultural Revolution – which occupied over 80% of the participants for the CR/10 project – they must have numerous memories of the historical period that need to be organized so as to be expressed. However, this process is deemed time consuming. With the ten-minute grammar, the project can save time with one particular interviewee and collect as many distinct memories as possible, so as to construct the multifaceted atlas of Cultural Revolution memories.

Serving as a basic function unit for the memory atlas, the ten-minute grammar generates tension between "remembering" and "forgetting." Browsing the video collection, one may frequently hear the interviewees say that they could not recall any details, only the "snippets." Some remembered only the sensational death scenes they encountered; some may recite the *Quotations from Chairman Mao* word by word but cannot recall any specific events. Some only talked about the happy moments they experienced during that period of time, as if Cultural Revolution never have posed any impacts. The ten-minute grammar forces the interviewees to focus on certain topics, while simultaneously choosing to leave others out. None of the ten minutes is complete by themselves; but collectively they can inform diversified understandings of the historical chapter. Combined with three pathways, the ten-minute grammar turns CR/10 into a memory atlas.

### *Catalogue: Randomization and an Atlas of Paradoxical Histories*

Catalogue can count as the first and foremost navigation pathway on CR/10. Similar to all digital archival collections, the catalogue of CR/10 website performs the basic function of preserving and identifying collected oral historical accounts. With the randomly assigned identifier to each video, a user can easily perform the task of search and locate a specific video. Besides the basic function, the identifier in CR/10's catalogue also serves as a *randomized index* of the fragmented Cultural Revolution memories. The *randomization* of the videos, in this case, functions not only as a gesture for an ethical, neutral treatment of interviewees and their life stories, but more importantly, as a mechanism that displays *paradoxical histories* of the Cultural Revolution and triggers emotional responses among users. Interview narratives organized in random orders may contradict with each other; a painful account on the brutality of Red Guards may be contrasted with a nostalgic recollection of the happy childhood during the same period. This design feature functions as a gentle warning message for the users when interacting with the CR/10 interface, to hold back from drawing generalizations solely based on the accounts they have encountered and to consider the various versions of the story. From the perspective of Warburg's *Mnemosyne Atlas*, randomization creates distance between each indexed object, and this in-between distance inspires both emotional responses and critical reflections among users, sustaining the idea of a multi-narrative memory atlas for China's Cultural Revolution.

### Timeline: Temporality of Memories and An Atlas of Forgetting

In addition to the catalogue, a timeline was created as one of the pathways in the interface to facilitate collection browsing and model the temporality of cultural memories on the Cultural Revolution. As Pierre Nora stated in his characterization of memories:

"[Memory] remains in permanent evolution, open to the dialectic of remembering and forgetting, unconscious of its successive deformations, vulnerable to manipulation and appropriation, susceptible to being long dormant and periodically revived. Memory is a perpetually actual phenomenon, a bond tying us to the eternal present."[44]

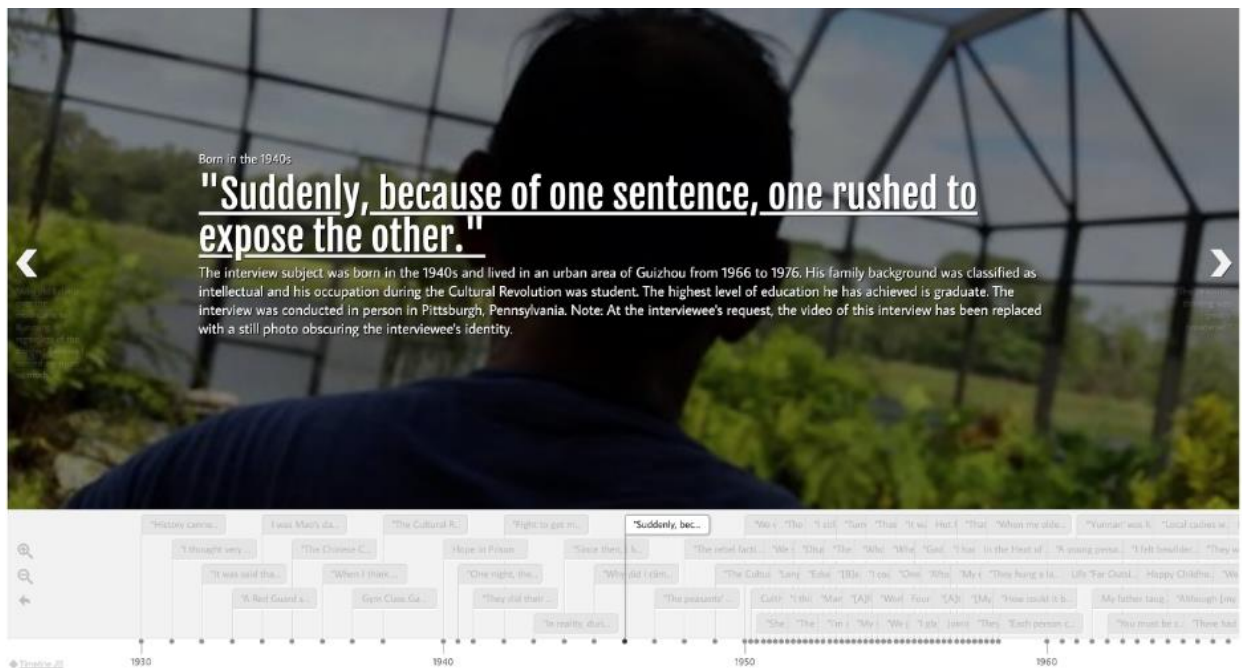

**Fig. 2.** Timeline on the CR/10 interface, created with TimelineJS (http://culturalrevolution.pitt.edu/).

Organized based on the time when interviewees were born, the timeline (see fig. 2) traces the chronological flow of memories across generations. Interviewees' birth period information was used to organize the timeline because it reflects the approximate age of the interviewee when they experienced the Cultural Revolution, which would presumably influence an individual's experiences and memories of the historical incident. Navigating through the timeline, one can easily observe the gradual yet inevitable process of oblivion, forgetting, and ignorance of the historical period. The first interview displayed on the timeline features a woman born in the 1910s, who was then a translation scholar and recalled her experiences of having all the possessions confiscated by the Red Guards and surviving in her basement for ten years. Majority of the video accounts gather around the 1950s and '60s and display the widest range of narratives about the historical incident. Moving into the 1970s, 1980s, and 1990s where the timeline concludes, only a limited number of 21 videos have been collected to highlight the impressions on the Cultural Revolution from younger generations. In the *atlas of forgetting* that is shaped around the time, young interviewees from both China and the United States discussed how they have learned *so little* about the historical incident from history textbooks, popular literary or cinematic works, or their families, due to various social and cultural reasons. One interviewee clearly stated that she has no interests in the Cultural Revolution because it is *so far away* from her life.[45]

### *Map: Spatiality of Memories and an Atlas of Geographical Impacts*

Along with the timeline, another pathway of CR/10 is the map (see fig. 3) that addresses the assumption: Do CR/10 memories vary across geographical locations? On this map, all the interview videos are organized based on the major geographical areas of the interviewee, which more specifically refers to the provinces they lived in during the Cultural Revolution. For younger interviewees who have not experienced the Cultural Revolution, the geolocation means the provinces they were born and raised in. Under the provincial category, each interview is further classified into either the "countryside" group or the "city area" group, basing off the interviewee's living experiences. With a click on a specific province, one can identify a complete list of interviews associated with this geographic area, grouped into the rural and urban categories. The darker red areas on the map showcase geographical regions that have not been represented so far in the video collection, such as Tibet, Xinjiang, Gansu, Fujian, and Taiwan. This may be attributed to the limitation of the snowball sampling method utilized during the interviewee recruiting process.[46] While it is early to conclude that Cultural Revolution had the least impacts on such regions, the map does show that historical impacts of the Cultural Revolution are not geographically even, with Beijing being the most heavily influenced region and the southern, remote, rural regions receiving the least impact. Life experiences, as the data map has demonstrated, also differ in central regions (e.g., Beijing) and southern areas (e.g., Guangzhou province).[47]

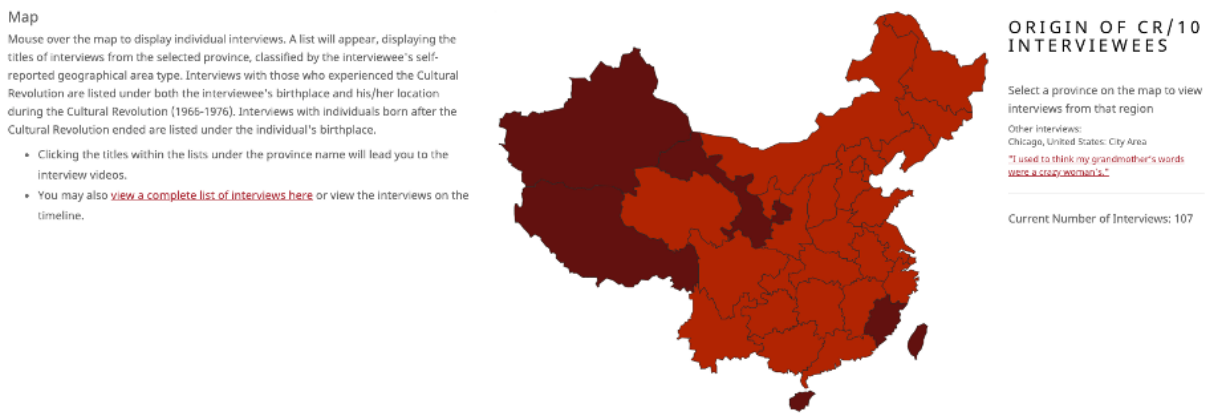

**Fig. 3.** Map on the CR/10 website (https://culturalrevolution.pitt.edu/).

The ten-minute grammar along with the three pathways shaped CR/10 interface as a Warburgian memory atlas that leverages distance to generate non-linear memory narratives. Such design features enable users to navigate the collection from various ways and explore the Cultural Revolution from diverse perspectives. While the CR/10 memory atlas does not implement one specific group narrative, it indeed makes an argument with the Warburgian design that China's Cultural Revolution is multiple and complex.

## CR/10 In Reception: User Experiences Studies

To further demonstrate how the design of CR/10 website as a memory atlas is received among users, I conducted a series of user experience interviews with individuals both within and outside academia, who demonstrated various degree of knowledge about China's Cultural Revolution. The focus of the user studies was to examine the usability of CR/10 website, especially the effects of the Warburgian design on shaping multifaceted narratives of the Cultural Revolution among users. By means of the user research, I aim to provide recommendations to further improve the design of CR/10.

### Data and Methods

For the user research, I identified two major user groups: One is the academic cohort, including China scholars, graduate students, and academic librarians, with specialized knowledge of China's Cultural Revolution. The other group is the general users, which in this case, includes K-12 teachers, school librarians, or other educational professionals with enthusiasm for China topics. The pilot study in 2018 focused on the academic cohort who demonstrate a relatively high level of knowledge of China's Cultural Revolution. I actively engaged in a roundtable on CR/10 during a major Asian Studies conference, where five China scholars from fields of literary studies, film and media studies, and histories discussed the academic values of CR/10 and their experiences using the CR/10 interface. Extending from the results, to further illustrate how CR/10 was received among general users, I performed a second round of research, which contained (1) a virtual ethnography with 25 book group participants, and (2) six semi-structured interviews with candidates selected from the book group. Table 1 shows an overview of the user research design, including three studies and their corresponding datasets and target users.

**Table 1.** Overview of User Research Design.

| Study | Data | User Group ID |
|---|---|---|
| Roundtable | Data memo 1 | S1 |
| | Data memo 2 | S2 |
| | Data memo 3 | S3 |
| | Data memo 4 | S4 |
| | Data memo 5 | S5 |
| Virtual Ethnography | Fieldnote 1 | Book group |
| | Fieldnote 2 | Book group |
| Semi-Structured Interview | Transcript 1 | G1 |
| | Transcript 2 | G2 |
| | Transcript 3 | G3 |
| | Transcript 4 | G4 |
| | Transcript 5 | G5 |
| | Transcript 6 | G6 |

For the virtual ethnography, I joined an online book group on "Mao and the Chinese Cultural Revolution," organized by the National Consortium for Teaching about Asia (NCTA) on ProBoards. ProBoards is a free message board service platform that facilitates online discussions. The NCTA Book Group is a semi-public forum that has recruited a number of 25 K-12 teachers, librarians, and other educational professionals interested in China and the Cultural Revolution period. The discussion materials included a biography of Mao Zedong by the Chinese historian Jonathan Spence and a number of recent digital archival materials including CR/10.

I conducted participatory observations in the online book group for three months from the beginning of the book group until its end.[48] During the process, I observed weekly online interactions among users while participating in the design of prompts for CR/10. Following the conclusion of the virtual ethnography, I selected six individuals from the book group to conduct further in-depth, semi-structured interviews based on their professional backgrounds and responses to CR/10 prompts. Among the six interviewees, five were high school teachers and their teaching subjects included world histories, political science, social studies, and English. Only one recruited interviewee was an elementary school librarian. All the interviewees live in the United States, except for one who was momentarily visit-teaching in China during the time of the interview. All interviews were conducted virtually through Skype and were audio-recorded in full for transcription and analytical purposes.

Each interview contained two major sets of questions: (1) Which pathway(s) do users prefer to navigate the CR/10 collection? (2) How would the users evaluate the pros and cons of each pathway, especially their usability and rhetorical effects? During the interview, each participant was asked to explore the CR/10 website independently for 15 minutes, trying out three navigation tools (i.e., the pathways). After the independent navigation, I worked with the interviewees to reflect on their experiences navigating through the collection.

### User Research Findings

Two stages of user experience research demonstrated distinct navigation preferences among the academics and general users. The academic cohort demonstrated to be more appreciative of the *complexity* and *flexibility* of Cultural Revolution narratives facilitated by the multiple pathways on CR/10, especially the timeline and the map. The pathways of timeline and map were evaluated by scholars as the features that demonstrate the most academic values of CR/10, as they represent a wide range of possibilities to shape Cultural Revolution narratives and the potential to model and test assumptions of the Cultural Revolution. They function, essentially, as what Johanna Drucker identified as the *humanistic* interface, where the infrastructure itself engages in making arguments.[49]

Scholars at the conference have also demonstrated interests in more complex navigational features to organize and retrieve videos from the CR/10 interface. For instance, one scholar indicated research intentions to "*only focus on interviews with women, so as to explore gendered expressions of the Cultural Revolution memories*" and to "*particularly look for children's accounts in the collection*".[50] Another scholar raised a potential topic on "taboo accounts" during the Cultural Revolution, one that particularly examines videos in which the informants had their images covered or voices distorted.[51] These research interest exemplify academic users' major demands for the increased searchability and analytical power of CR/10.

In comparison, general users with a lower level of familiarity with Cultural Revolution appeared to have a stronger preference for the conventional catalogue method. Two teachers and the librarian I interviewed said they preferred to use the video titles to navigate the collection. The title of a video in CR/10 is usually a quotation from the original video that highlights its most essential message. As one participant claimed in the interview,

> "I browsed the *quotations* to find interviews that I felt interested in. … If I were to use interviews for teaching, I might use them based on the age of the interviewees or where they were from in China. But for my purposes, I just pulled them out by what quote interested me the most…and that's how I found them."[52]

Another interviewee also identified catalogue as his most preferred method of navigation, saying that titles "give you an idea of what the main point of the video was."[53] This contrasting preference between scholars and non-academics may be simply explained by the fact that general users are not necessarily equipped with the background knowledge to interpret messages from the timeline or the data map of the Cultural Revolution. As one user indicated,

"I thought the map was interesting. But because regions do not mean as much to me as they mean to other people – I mean those who know the regions very well, I would like to approach it (i.e., the collection) with other ways."[54]

Beyond the navigation preference, general users also indicated stronger curiosity for individual life experiences, rather than a systematic interpretation of memory narratives. When asked what improvements they could think of for CR/10, one user immediately stated that,

"it would be interesting to give updates on those interviewees, like where are they now and what are they doing? Kind of like their current status. Some [of the video interviewees] did talk a little bit about that. But it might help people place them in terms of where they ended up, what all of that have brought them to, and where they are now."

As can be seen from the discussions above, the user studies suggest different degrees of usability for the three pathways. The randomized catalogue proves to be the most effective for general, non-academic users to browse the collection and develop a basic understanding of the Cultural Revolution by watching fragmented yet compelling individual stories of the Cultural Revolution. The timeline and map, by contrast, demonstrate to have stronger analytical power to facilitate systematic memory narratives, but require the users to have certain historical knowledge to effectively interact with them. All the three pathways, however, have effectively invited users to deeply engage with the collection, either by attracting them with paradoxical individual memory accounts or the analytical potential of design features.

***Implications of the Results***

Results from the user studies inform multiple new design opportunities for CR/10. The first and foremost opportunity, as illustrated in the user research with the scholar cohort, is to increase the searchability of the interface. On the current CR/10 website, the timeline and map features support only browsing, rather than searching. Catalogue is the only method for searching, but the searchable information and categories are limited to those included in metadata. Enabling search functions in the interface, however, would more directly encourage users to interact with the memory data. Related to the increase in searchability, constant updates on the searchable information of videos would also be important to satisfy the user needs such as to find gendered expressions or the "taboo accounts" of the Cultural Revolution. Scholars have also demonstrated the interests in performing analytical tasks within the interface and exploring "live results."[55] Infrastructural works of this kind in the area of Asian research include only a few examples, such as the China Biographical Database Project (CBDB)[56] at Harvard University and the Chinese Text Project[57], where analytical tools are embedded in the interface to empower real-time data analyses. Such a consideration, as discussed earlier in this article, also corresponds to the increasing computational trends in the fields of digital archives, despite the fact that it requires a higher level of technological and institutional support for implementation. The final opportunity at this point is aligned with the interests of general users to interact with individual accounts and personal memories. From this perspective, CR/10 can also benefit from encouraging more public input. To further engage the general users in the continuous evolvement of the Cultural Revolution memories, enabling an "open feedback" field would be instrumental to encourage participation from the public, to either highlight messages from videos, or share their thoughts and comments on the interface. Applications of this crowdsourcing model in digital humanities has also been acknowledged in scholarly works.[58]

### Challenges and Limitations of the User Studies

The user research also suffers from few limitations. First, as the CR/10 website is not accessible from the mainland China, the user studies have only focused on audiences in the United States, and particularly professionals working in education and research enterprises. Although this intentional choice was made based on the initial orientation of CR/10 as a research and educational platform, studies with users from more diverse domain knowledge and personal interests will be beneficial to further improve the usability as well as the design of CR/10. In addition to the user groups, the number of data sample was another limitation of the user research. Although the semi-structured interviews and ethnographic research identified important issues of usability for CR/10 and were able to inform future design possibilities, a survey with a larger user pool will be particularly useful to find the most valuable option or the optimal plan for future implementation.

## Conclusion

As the infrastructural building advances rapidly for digital humanities research, digital archives and digitalized cultural records undoubtedly serve important roles in shaping humanities knowledge and research practices.[59] However, as new paradigms just began to take form, what might be *effective infrastructure* for the humanities research remains a question. This paper concerns this broad question with a specific case of CR/10, an interactive digital oral history platform designed to engage the public with memory works on China's Cultural Revolution. Presenting a rhetorical analysis of the major design features and a series of user studies, I demonstrated in this article that CR/10 interface functions as a Warburgian memory atlas and has successfully given rise to multifaceted memory narratives among different groups of audiences. By identifying CR/10's major user groups and their respective needs for interaction with the interface, I proposed design recommendations from three aspects, which essentially include (1) increasing searchability, (2) empowering computational and analytical functions, and (3) enabling open feedback from the general public. Such design efforts may have the potential to further shape CR/10 into a public-oriented, digital humanities interface, in addition to an ever-moving atlas of memories.

**Author Declaration:** This work was not supported by any grant and does not have any conflict of interest.

## Notes

[1] "Decision of the CC of the CPC Concerning the Great Proletarian Cultural Revolution," https://www.marxists.org/subject/china/peking-review/1966/PR1966-33g.htm.

[2] East Asian Library ULS, University of Pittsburgh, "CR10," 2016, https://pitt.hosted.panopto.com/Panopto/Pages/Viewer.aspx?id=a82a0c82-9dc5-464e-aa0b-4871a3d08f68.

[3] Sandi Ward, "Collecting Oral History in an Academic Library: The CR/10 Project," *Journal of East Asian Libraries*, no. 167 (2018): 8.

[4] Cornell University Library, "Mnemosyne," https://warburg.library.cornell.edu/.

[30] Aby Warburg completed his dissertation in 1891, titled *Botticelli's "Birth of Venus" and "Spring." An Examination of the Representations of Antiquity in the Early Italian Renaissance.* His research laid a basis for what was later called the "iconographic methods" in art history. This method was further developed by Warburg's student, Erwin Panofsky, who authored the *Studies in Iconology.* Matthewa Rampley, "Iconology of the Interval: Aby Warburg's Legacy," *Word & Image* 17, no. 4 (October 2001): 303–24.

[31] Martin Warnke and Lisa Dieckmann, "Prometheus Meets M Eta-Image: Implementations of Aby Warburg's Methodical Approach in the Digital Era," *Visual Studies* 31, no. 2 (June 2016): 109–20.

[32] Charlotta Krispinsson, "Aby Warburg's Legacy and the Concept of Image Vehicles. 'Bilderfahrzeuge': On the Migration of Images, Forms and Ideas. London 13-14 March 2015.," *Konsthistorisk Tidskrift/Journal of Art History* 84, no. 4 (October 2, 2015): 244–47.

[33] Christopher D. Johnson, *Memory, Metaphor, and Aby Warburg's Atlas of Images*, Signale: Modern German Letters, Cultures, and Thought (Ithaca, N.Y: Cornell University Press: Cornell University Library, 2012).

[34] There exists a large number of scholarly interpretations for the panels in *Mnemosyne Atlas*, and a close review of each piece of work goes far beyond the scope of this paper. More bibliographical details can be found in Cornell University Library, "Mnemosyne."

[35] Warburg's own introductions to the *Mnemosyne Atlas*, quoteed from Rampley, "Iconology of the Interval."

[36] David L. Marshall, "Warburgian Maxims for Visual Rhetoric," *Rhetoric Society Quarterly* 48, no. 4 (August 8, 2018): 352–79, https://doi.org/10.1080/02773945.2017.1411602.

[37] M. Lynch, "The Externalized Retina: Selection and Mathematization in the Visual Documentation of Objects in the Life Sciences," in *Representation in Scientific Practice*, ed. Michael Lynch and Steve Woolgar (Cambridge, Massachusetts: The MIT Press, 1990).

[38] Scott Montgomery, *The Chicago Guide to Communicating Science* (University of Chicago Press, 2002).

[39] Kurt W. Forster, "Aby Warburg's History of Art: Collective Memory and the Social Mediation of Images," *Daedalus* 105, no. 1 (1976): 169–76.

[40] Rampley, "Iconology of the Interval."

[41] Warburg quoted in Warnke and Dieckmann, "Prometheus Meets M Eta-Image."

[42] Ibid.

[43] Walter Schneider and Richard M. Shiffrin, "Controlled and Automatic Human Information Processing: I. Detection, Search, and Attention.," *Psychological Review* 84, no. 1 (1977): 1.

[44] Pierre Nora, "Between Memory and History: Les Lieux de Mémoire," *Representations* 26 (April 1, 1989): 7–24.

[45] East Asian Library ULS, University of Pittsburgh, "'I'm Not Interested.' (CR10-0045-HLJ)," CR/10, accessed January 14, 2021, https://digital.library.pitt.edu/islandora/object/pitt%3A7198631/viewer.

[46] Snowball sampling is a chain referral sampling technique to recruit participants, where the researcher starts from their personal networks to recruit the first participant and then expand the recruitment based on referrals of the participants. Problems of this technique for qualitative research are discussed in Patrick Biernacki and Dan Waldorf, "Snowball Sampling: Problems and Techniques of Chain Referral Sampling," *Sociological Methods & Research* 10, no. 2 (1981): 141–63.